\documentclass[aps,prb,preprint,groupedaddress]{revtex4}

\usepackage{graphicx}
\usepackage{amssymb}
\usepackage{bm}
\usepackage{graphpap}
\begin{document}

\title{Excitons in Silicon Nanocrystallites: the Nature of Luminescence}

\author{Eleonora \surname{Luppi}, Federico \surname{Iori}, Rita
\surname{Magri}} \affiliation{CNR-INFM-$S{^3}$ and Dipartimento di
Fisica - Universit\`a di Modena e Reggio Emilia - via Campi 213/A
I-41100 Modena Italy.}
\author{Olivia  \surname{Pulci}}
\affiliation{European Theoretical Spectroscopy Facility (ETSF) and
CNR-INFM, Dept. of Physics, Universit\'a di Roma "Tor Vergata" Via
della Ricerca Scientifica 1, I-00133 Roma, Italy}
\author{Stefano \surname{Ossicini}, Elena \surname{Degoli}}
\affiliation{CNR-INFM-$S{^3}$ and Dipartimento di Scienze e Metodi
dell'Ingegneria - Universit\`a di Modena e Reggio Emilia - via
Amendola 2 Pad. Morselli, I-42100 Reggio Emilia Italy.}
\author{Valerio \surname{Olevano}}
\affiliation{European Theoretical Spectroscopy Facility (ETSF) and
Laboratoire d'\'Etudes des Propri\'et\'es \'Electroniques des
Solides,
 UPR 11, CNRS, F-38042, Grenoble, France}
\date{\today}

\begin{abstract}
The absorption and emission spectra of silicon nanocrystals up to
1 nm diameter including geometry optimization and the many-body
effects induced by the creation of an electron-hole pair have been
calculated within a first-principles framework. Starting from
hydrogenated silicon clusters of different size, different Si/O
bonding at the cluster surface have been considered. We found that
the presence of a Si-O-Si bridge bond originates significative
excitonic luminescence features in the visible range that are in
fair agreement with the experimental outcomes.
\end{abstract}
\pacs{73.22-f; 71.24.+q,73.20.at; 78.67.Bf.}
 \maketitle

Following the initial impulse given by the discovery of
photoluminescence (PL) from porous silicon \cite{can},
nanostructured silicon has received extensive attention
\cite{bisi,ossospringer}. This activity is mainly focused on the
possibility of  getting relevant optoelectronic properties from Si
nanocrystals (Si-nc). Optical gain, observed in Si-nc embedded in
SiO$_2$, has given further impulse to these studies \cite{pav}. It
is generally accepted that the quantum confinement (QC), caused by
the nanometric size, is essential for the PL in Si-nc, but the
interpretation of the PL features, such as the substantial
redshift (RS) of the PL energy with respect to the theoretical
predictions based merely on the QC model and its independence from
the size for small ($< 3$ nm) crystallites, is still
controversial. Baierle \emph{et al.} \cite{bai} and G. Allan
\emph{et al}. \cite{allan} stressed the importance of bond
distortion at the Si-nc surface in the excited state (ES) in
creating an intrinsic localized state responsible of the PL
emission. Wolkin \emph{et al}. \cite{wol} observed that also
oxidation introduces states in the gap, which pin the transition
energies. They and others \cite{gal,lup1} suggest that the
formation of a Si=O double bond is responsible of the RS upon
oxidation of the optical absorption edge.  On the contrary,
Vasiliev \emph{et al.} \cite{vas} showed that absorption gaps of
similar size can be obtained also for O connecting two Si atoms
(bridge bond) at the Si-nc surface. Recently Gatti and Onida
\cite{gatti} considered six small different prototypical oxidized
Si clusters and found results similar to those of \cite{gal,lup1},
i.e. that the RS of the absorption edge is much more pronounced in
the case of the double Si=O bond that for bridge bonds. Ramos et
al. \cite{ramos} have found a blueshift of the absorption onset
for strongly oxidized Si-nc. Actually, heavy oxidation (the
formation of complete oxide shells) originates a reduction of the
effective size of the Si-nc, i.e. an increase of QC. Although all
these calculations address only the problem of absorption, yet the
large majority of the experimental results and the most
interesting ones are relative to PL measurements, thus are
strictly related to excited state results. To date very few papers
have addressed the issue of the ES configurations, which is mostly
relevant for Si-nc with a high surface to volume ratio.
Theoretically their description has been performed using the so
called $\Delta$SCF method \cite{gius,puzder,franceschetti,degoli},
where total energies are calculated both in the ground (GS) and
excited states (ES). Here the ES corresponds to the electronic
configuration in which the highest occupied single-particle state
(HOMO) contains a hole ($h$), while the lowest unoccupied
single-particle state (LUMO) contains the corresponding electron
($e$). Thus, one can extract the absorption and emission energies
and through their difference calculate the Stokes or Frank-Condon
Shift due to the lattice relaxation induced by the electronic
excitation. The obtained results show a dependence of the Stokes
Shift (SS) on the Si-nc size, that is less marked for the O double
bonded case.\\ Our aim, here, is to allow a direct comparison
between experimental data and theoretical results, thus we
calculate not only the transition energies but also directly the
absorption and emission optical spectra. For both the GS and ES
optimized geometry, we have evaluated the optical response $Im~
[\varepsilon_{NC}(\omega)]$ (the imaginary part of the nanocrystal
dielectric function) through first-principle calculations also
beyond the one-particle approach. We consider the self-energy
corrections \cite{Hedin} by means of the GW method and the
excitonic effects through the solution of the Bethe-Salpeter (BS)
equation \cite{reviewOnida}. The effect of local fields (LF) is
also included, to take into account the inhomogeneity of the
systems. In this work, the emission spectrum has been calculated,
in a first approximation, as the time reversal of the absorption
\cite{bassani}. Strictly speaking, $Im~[\varepsilon_{NC}(\omega)]$
corresponds to an absorption spectrum in a new structural
geometry, the ES geometry, with the electronic configuration of
the GS. For the first time, the electron-hole interaction is here
considered also in the emission geometry. This different approach
where the many-body effects
 are combined with the study
of the structural bond distortion at the Si-nc surface in the ES,
account both for the observed absorption and PL spectra. The
procedure is exemplified for Si-nc of different size, a small
Si$_{10}$H$_{16}$ (0.55 nm diameter) and a larger
Si$_{29}$H$_{36}$ clusters (0.9 nm diameter). We demonstrate that
light emission in the near-visible range of Si-nc is related to
the presence at the Si-nc surface of Si-O-Si bridge bonds.
Actually we find that a strong excitonic peak at about 1.5 eV
emission energy is obtained in the Si$_{10}$H$_{16}$ cluster in
the presence of a Si-O-Si bridge bond as result of a considerable
bond distortion on the electron-hole interaction, while the
emission corresponding to the Si=O double bond is predicted to
occur at a much larger energy as a consequence of a much smaller
structural distortion and of a first electronic transition almost
dark. Similar results have been obtained for the larger
Si$_{29}$H$_{36}$ cluster.

\begin{figure}[b]
\includegraphics[clip,width=0.75\textwidth]{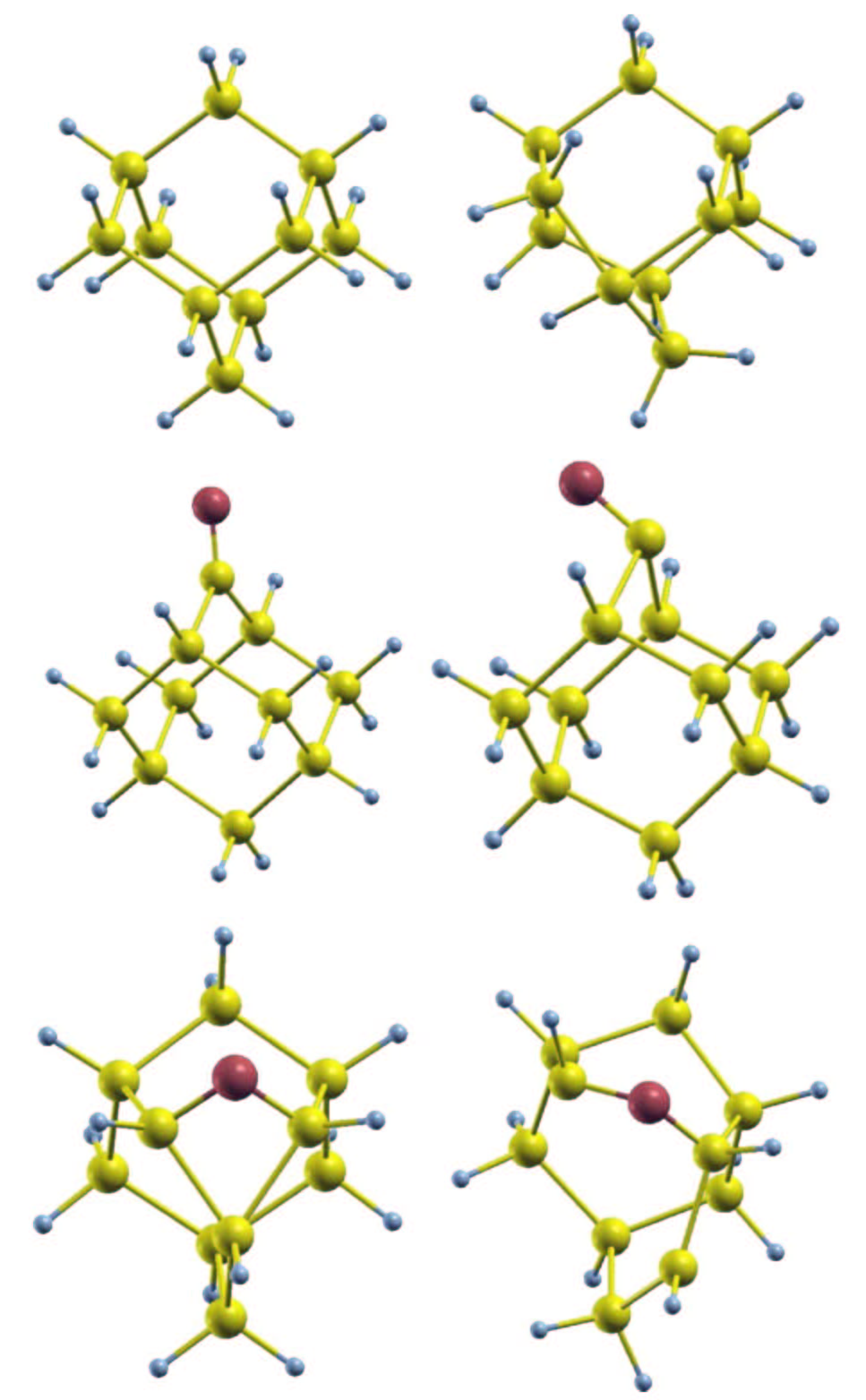}
\caption{\small (Color online) Calculated structures for the
Si$_{10}$H$_{16}$ (top panel), Si$_{10}$H$_{14}$=O (central panel)
and Si$_{10}$H$_{14}>$O (bottom panel) clusters at relaxed
geometry in both the ground- (left panels) and excited-state
(right panels). Grey (yellow) balls represent Si atoms, the black
(red) ball is the O atom, while the small grey (grey) balls are
the hydrogens used to saturate the dangling bonds.} \label{strutt}
\end{figure}

The starting configuration for all clusters is the hydrogenated
structure (the Si$_{10}$H$_{16}$ and the Si$_{29}$H$_{36}$ ones);
next, we consider two types of Si/O bonds at the clusters surface:
the silanone-like Si=O bond (Si$_{10}$H$_{14}$=O and
Si$_{29}$H$_{34}$=O) and the Si-O-Si bridge bond, where the O atom
is in between two Si atoms as in the SiO$_2$ (Si$_{10}$H$_{14}>$O
and Si$_{29}$H$_{34}>$O). It is worth to note that this type of
bridge bond has been demonstrated to lead to the stablest isomer
configuration by Gatti and Onida \cite{gatti,sym}.  Full
relaxation with respect to the atomic positions is performed for
all systems both in the ground and excited configurations
\cite{pwscf}. Figure~\ref{strutt} presents the relaxed optimized
structures of the considered Si$_{10}$ based clusters in their
electronic GS and ES configuration. The ionic relaxation has
produced structural changes with respect to the initial geometry
which strongly depend on the type of surface termination.

\begin{table}[hb]
\begin{center}
\begin{small}
\caption{Absorption and emission optical gaps calculated within
DFT-LDA, GW, and with the inclusion of excitonic and local field
effects (BS-LF).
In parenthesis also the lowest dark transitions (when present) are
given. All values are in eV. }\label{gaps}
\begin{tabular}{r | c c }
  & \multicolumn{1}{c}{\textbf{Absorption}}
  &{\textbf{Emission}}\\
    & \multicolumn{1}{c}{\textbf{LDA GW BS-LF}}
  &{\textbf{LDA GW BS-LF}}\\
\hline
\textbf{Si$_{10}$H$_{16}$} & 4.6, 8.6, 5.2  & 0.1, 3.8, 0.4\\
\textbf{Si$_{10}$H$_{14}=$O} &  3.3 (2.5),  7.3 (6.5),  3.7 (2.7) & 0.8, 4.6,  1.0 \\
\textbf{Si$_{10}$H$_{14}>$O} & 3.4, 7.6, 4.0 &0.1, 3.5, 1.5\\
\textbf{Si$_{29}$H$_{34}=$O} &  2.5,  6.0,  3.7 (3.1)  & 0.9, 4.1,  1.2 \\
\textbf{Si$_{29}$H$_{34}>$O} & 2.3, 4.8, 2.3 & 0.4, 3.0, 2.2 (0.3) \\
\end{tabular}
\end{small}
\end{center}
\end{table}

In Si$_{10}$H$_{16}$, some distortions occur at the surface in the
excited state as compared to the ground state geometry, and the
Si-Si distances in the core shells are somewhat concerned by the
excitation. In the case of Si$_{10}$H$_{14}$=O, the changes are
mainly localized near the O atom, in particular the angle between
the double bonded O and its linked Si atom is modified (see
Fig.~\ref{strutt}). In the bridge structure (Si$_{10}$H$_{14}>$O),
instead, the deformation is localized around the Si-O-Si bond
determining a  considerable strain in the Si-Si dimer distances
\cite{sym}. The outcomes are similar for the larger Si$_{29}$ based
clusters.

The only
difference is that now the distortion induced by the promotion of an
electron in the excited state is smaller, as expected,
since for large clusters the charge density perturbation is
distributed throughout all the structure, and the effect locally
induced becomes less evident.

These structural changes are reflected in the electronic and optical
properties. This is shown in Table~\ref{gaps}, where the calculated
optical gaps (energy differences between LUMO and HOMO) at different
levels of approximation are reported for both the Si$_{10}$ and
Si$_{29}$ based nanocrystals. Concerning the transition energy
values of Table \ref{gaps} we see that, going from LDA to GW,  the
main results common to the absorption and emission cases are the
opening of the band-gap by amounts weakly dependent on the surface
termination but much larger than the corresponding 0.6 eV Si bulk
case. Looking at the BS-LF calculations, we note a sort of
compensation (more evident in the GS than in the ES) of self-energy
and excitonic contributions: the BS-LF values return similar to the
LDA ones. This compensation has been predicted theoretically by
Delerue et al. \cite{del} in zero-dimensional nano-materials. The
only exception are the BS-LF calculations for the excited state
geometries of the clusters in the presence of the Si-O-Si
bridge bonds. \\

\begin{figure}[b]
\includegraphics[clip,width=0.75\textwidth]{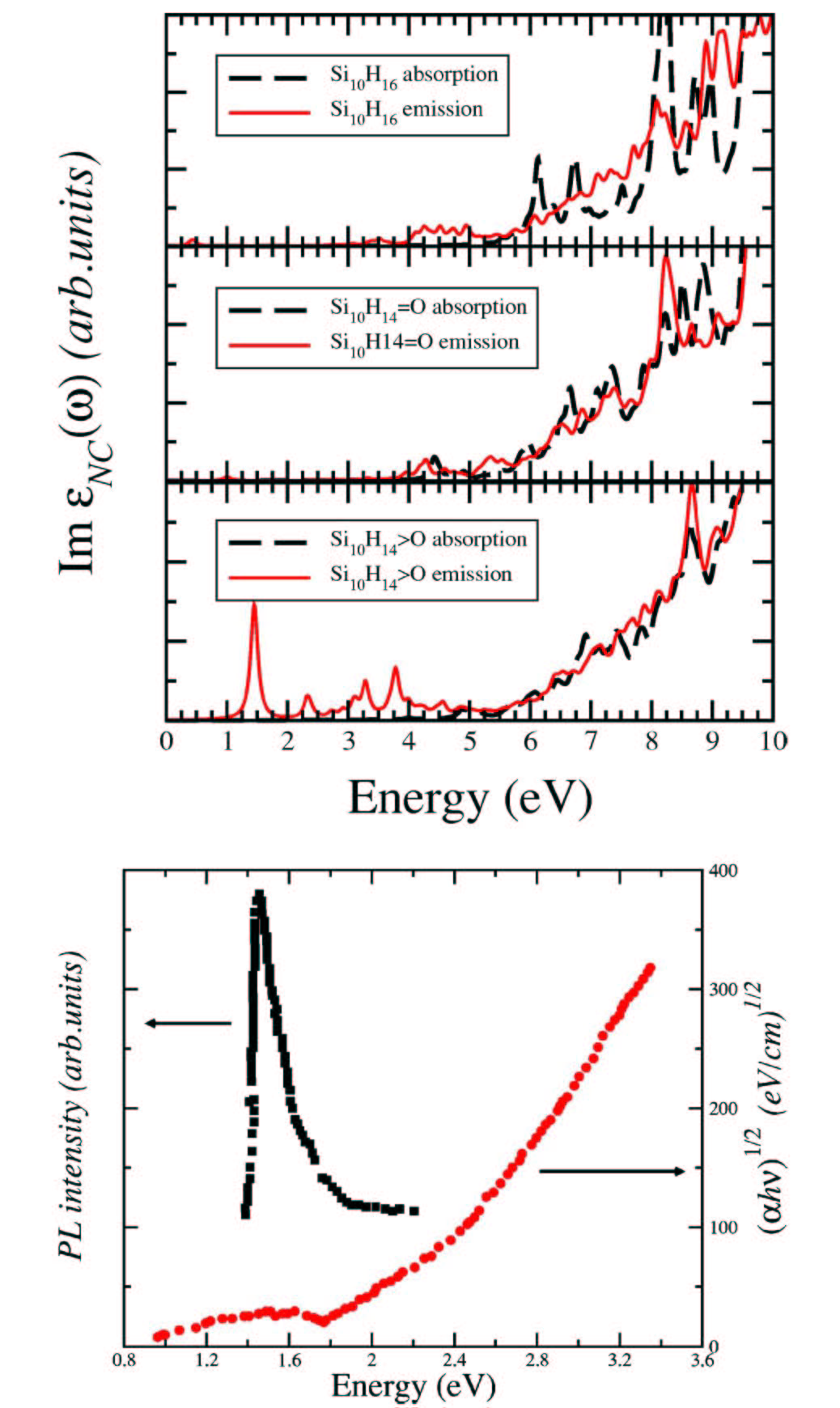}
\caption{\small Up: Emission (solid (red) line) and absorption
(dashed (black) line) spectra: imaginary part of the dielectric
function for the three considered Si-nc. Si$_{10}$H$_{16}$ (top
panel), Si$_{10}$H$_{14}$=O (central panel) and
Si$_{10}$H$_{14}>$O (bottom panel). Down: experimental results for
emission (left) and absorption (right) by \protect\cite{Ma}.}
\label{abs}
\end{figure}

Concerning the absolute values of all the absorption and emission
gaps it is worthwhile to note that whereas absorption predicts the
correct trend as function of size (i.e. larger gap for smaller
Si-nc) this is not the case of emission, where the situation is more
complex. This is due to the significant cluster distortion present
in the ES of the smaller Si-nc. Here the Stokes Shift is so strong
that the emission energies are now smaller for the smaller cluster
\cite{size}. This fact has been previously discussed for the case of
fully hydrogenated Si-nc, where, by total energy calculations
\cite{puzder,franceschetti,degoli}, it has been demonstrated that
the usual trend is recovered also for the emission for Si-nc with
diameter $\geq$ 1 nm. Having the absorption and emission gaps, we
can now give an estimate of the Stokes shift fully including
excitonic effects (the BS-LF results). Going from the ground to the
excited state geometry for the Si$_{10}$ based clusters, the fully
hydrogenated clusters show a remarkable Stokes shift (4.8 eV)
whereas the double bonded O clusters and the bridge bonded O
clusters present practically the same shift (2.7 eV and 2.5 eV).
These last shifts become, in the case of the Si$_{29}$ based
clusters, 2.5 eV and 0.1 eV respectively. It should be noted that if
the SS were calculated simply as differences of emission and
absorption energies of the lowest transitions (without looking at
the oscillator strength, i.e. without considering if these
transitions are dark or not) for the Si$_{10}$H$_{14}$=O cluster we
would obtain a weaker SS (1.7 eV) and for the Si$_{29}$H$_{34}>$O
cluster a stronger SS (2.0 eV) \cite{size}. A clearer insight on
these results is offered by Fig. \ref{abs}, where the calculated
absorption and emission spectra for all the Si$_{10}$ based clusters
are depicted.

Self-energy, local-field and excitonic effects (BS-LF) are fully
taken into account. Concerning the absorption spectra (Fig.
\ref{abs}, dashed lines), all three cases show a similar smooth
increase in the absorption features. Different is the situation for
the emission related spectra (Fig. \ref{abs}, solid lines). Here,
whereas the situation remain similar for the fully hydrogenated
Si$_{10}$H$_{16}$ (top panel) cluster and for the
Si$_{10}$H$_{14}$=O (central panel) cluster, in the case of a
Si-O-Si bridge bond (Fig. \ref{abs} (bottom panel)) an important
excitonic peak, separated from the rest of the spectrum, is evident
at 1.5 eV. Actually  bound excitons are present also in the fully
hydrogenated (at 0.4 eV) and in the Si$_{10}$H$_{14}$=O (at 1.0 eV)
clusters, with calculated binding energies (the energy difference
between the GW and the BS-LF results) even larger than in the case
of the Si-O-Si bridge bond (3.4 and 3.6 eV respectively, to be
compared with a binding energy of 2.0 eV in the case of the bridge
bond cluster). Nevertheless, the related transitions are almost dark
and the emission intensity is very low. Only in the case of the
Si-O-Si bridge bond a clear PL peak appears thanks to the strong
oscillator strength of the related transition. The bottom of Fig.
\ref{abs} shows the experimental absorption and emission spectra
measured by Ma et al. \cite{Ma} for Si-nanodots embedded in SiO$_2$
matrix. A strong photoluminescence peak appears around 1.5 eV.

\begin{figure}[b]
\includegraphics[clip,width=0.75\textwidth]{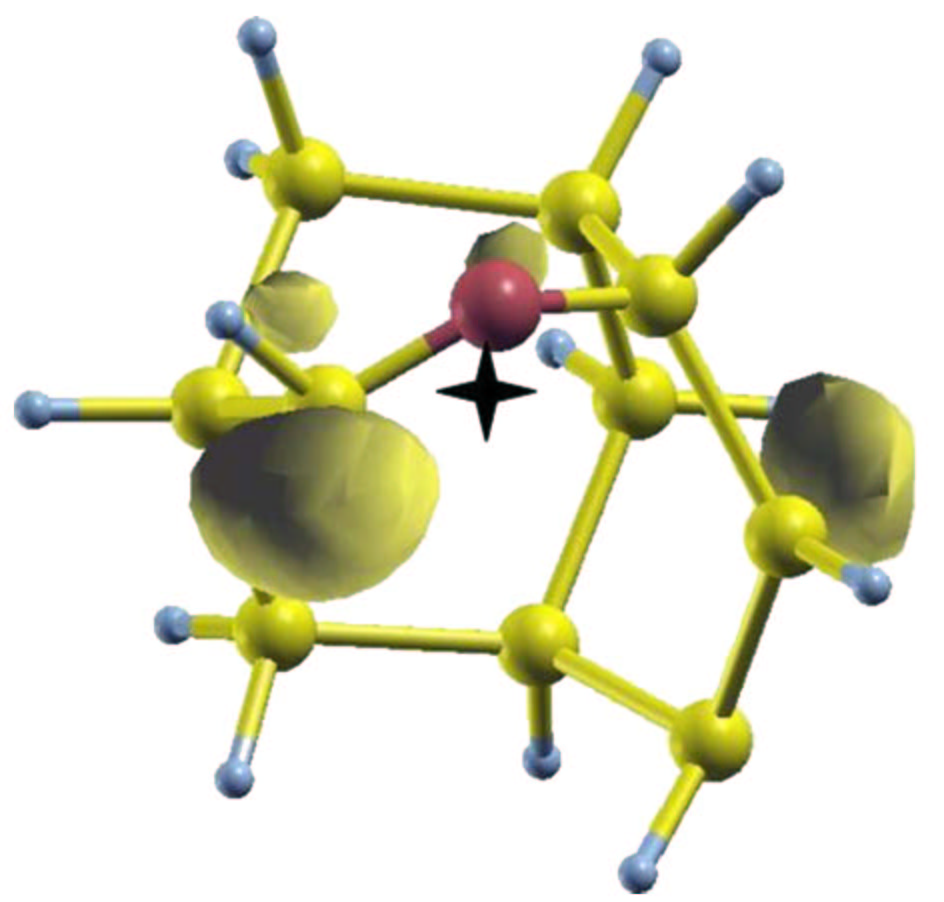}
\caption{\small Geometrical structure of the Si$_{10}$H$_{14}>$O
in the excited state configuration. Grey (yellow) balls represent
Si atoms, the dark grey (red) ball is the O atom, while the small
grey (grey) balls are the hydrogens used to saturate the dangling
bonds. The grey (yellow grey) isosurface give the probability
distribution $|\psi_{exc}(r_e,r_h)|^2$ for finding the electron
when the hole is fixed in a given position, represented by the
dark cross} \label{exc}
\end{figure}

\begin{figure}[b]
\includegraphics[clip,width=0.75\textwidth]{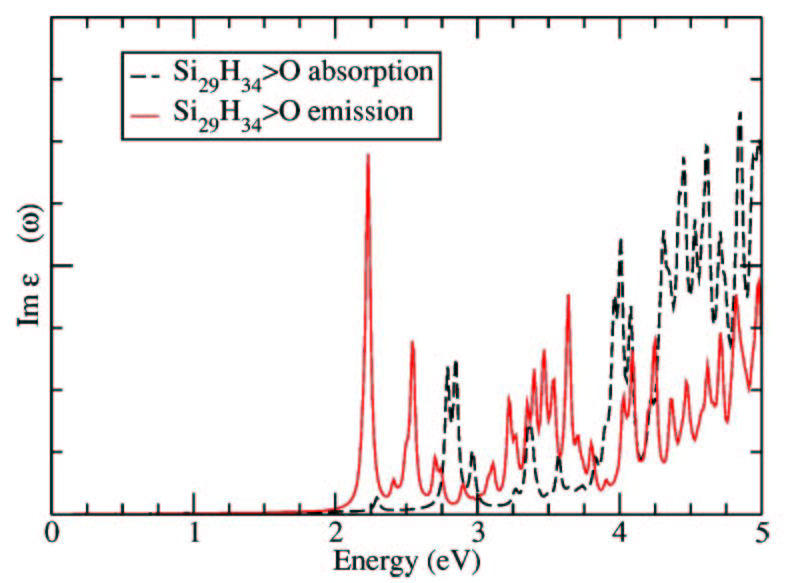}
\caption{\small Emission (solid line) and absorption (dashed line)
spectra: imaginary part of the dielectric function for the
Si$_{29}$H$_{34}>$O cluster.} \label{abs1}
\end{figure}

Comparison of the experimental spectra with our results suggest that
the presence of a Si-O-Si bridge bond at the surface of Si-nc and
the relative deformation localized around the Si-O-Si bond can
explain the nature of luminescence in Si nanocrystallites: only in
this case the presence of an excitonic peak in the emission related
spectra, red shifted with respect to the absorption onset, provides
an explanation for both the observed Stokes Shift and the
near-visible PL in Si-nc. These conclusions are supported by Fig.
\ref{exc} which shows  the real-space probability distribution
$|\psi_{exc}(r_e,r_h)|^2$ for the bound exciton as a function of the
electron position $r_e$ when the hole is fixed in a given position
$r_h$ (the dark cross in the figure). We see that the bound exciton
is mainly localized around the cage distortion. Similar conclusions
can be reached for the larger Si$_{29}$ nanocrystals. Fig.
\ref{abs1} shows the calculated absorption and emission spectra for
the Si$_{29}$H$_{34}>$O cluster, where the O atom is placed in a
bridge position as in the Si$_{10}$H$_{14}>$O case. Also in this
case starting from the Si$_{29}$H$_{36}$ cluster only in the case of
O in bridge position there is a cage distortion at the interface
that allows the presence of significant emission features in the
optical region. It is worthwhile to stress that the role of the
interface has been experimentally proven to be important for the PL
properties of embedded Si-nc in SiO$_2$ \cite{daldo} and in the
mechanism of population inversion at the origin of the optical gain
\cite{dalne,daldo}; besides, Monte Carlo approaches have
demonstrated that Si-O-Si bridge bonds are the main building blocks
in the formation of Si-SiO$_2$ flat interfaces \cite{tersoff} and
form the low energy geometries at the interface for Si-nc embedded
in silicon dioxide \cite{kelires}.

In conclusion, our theoretical results, obtained by ab-initio
calculations and fully including excitonic effects, suggest that the
Si-O-Si bridge bond is responsible for the strong PL peak
experimentally observed, and shed some light on the role of
Si-nc-$SiO_2$ interface.

This work is supported by MIUR (NANOSIM and PRIN 2005),
  by CNISM, and by the EU through
the NANOQUANTA Network of Excellence  (Contract No.
NMP4-CT-2004-500198). We acknowledge CINECA CPU time granted by
INFM (Progetto Calcolo Parallelo).
Bethe-Salpeter calculations have been performed using the EXC
code http://www.bethe-salpeter.org/.

\vfill\eject ..
\vskip 2.truecm
\vfill\eject
..
\vskip 2.truecm
\vfill\eject ..
\vskip 2.truecm
\vfill\eject ..
\vskip 2.truecm

\begin{thebibliography}{}
\bibitem{can} L. T. Canham, Appl. Phys. Lett. {\bf 57}, 1046 (1990).
\bibitem{bisi} O. Bisi, S. Ossicini, and L. Pavesi, Surf. Sci. Reports {\bf 38}, 5 (2000).
\bibitem{ossospringer} S. Ossicini, L. Pavesi, F. Priolo, "Light Emitting Silicon for Microphotonics",
Springer Tracts on Modern Physics {\bf 194}, Springer-Verlag
Berlin (2003).
\bibitem{pav} L. Pavesi, L. Dal Negro, C. Mazzoleni, G. Franz\'o, F. Priolo,
Nature {\bf 408}, 440 (2000); L. Dal Negro, M. Cazzanelli, L.
Pavesi, S. Ossicini, D. Pacifici, G. Franz\`o, F. Priolo, and F.
Iacona, Appl. Phys. Lett. {\bf 82}, 4636 (2003).
\bibitem{bai} R. J. Baierle, M. J. Caldas, E. Molinari, and S. Ossicini, Solid State Commun. {\bf 102}, 545
(1997).
\bibitem{allan} G. Allan, C. Delerue, and M. Lannoo, Phys. Rev. Lett. {\bf 76}, 2961 (1996).
\bibitem{wol} M. V. Wolkin, J. Jorne, P. M. Fauchet, G. Allan, and C Delerue, Phys. Rev. Lett. {\bf 82}, 197 (1999).
\bibitem{gal} A. Puzder, A. J. Williamson, J. C. Grossman, and G. Galli, Phys. Rev. Lett. {\bf 88}, 097401
(2002).
\bibitem{lup1} M. Luppi, and S. Ossicini,  J. Appl. Phys., {\bf 94}, 2130 (2003), Phys.
Rev. B {\bf 71}, 035340 (2005).
\bibitem{vas} I. Vasiliev, J. R. Chelikowsky, and R.M. Martin,
Phys. Rev. B {\bf 65}, 121302(R) (2002).
\bibitem{gatti} M. Gatti, G. Onida, Phys. Rev. B {\bf 72}, 045442
(2005).
\bibitem{ramos} L. Ramos, J. Furthm\"uller, F. Bechstedt, Appl.
Phys. Lett. {\bf 87}, 143113 (2005).
\bibitem{gius} E. Luppi, E. Degoli, G. Cantele, S. Ossicini, R.
Magri, D. Ninno, O. Bisi, O. Pulci, G. Onida, M. Gatti, A. Incze,
R. Del Sole, Opt. Mater. {\bf 27}, 1008 (2005).
\bibitem{puzder} A. Puzder, A. J. Williamson, J. C. Grossman, G. Galli,
J. Am. Chem. Soc. {\bf 125}, 2786 (2003).
\bibitem{franceschetti} A. Franceschetti, S. T. Pantelides, Phys. Rev. B {\bf 68},
033313 (2003).
\bibitem {degoli}  E. Degoli, G. Cantele, E. Luppi, R. Magri, D. Ninno, O. Bisi,
S. Ossicini, Phys. Rev. B {\bf 69}, 155411 (2004).
\bibitem{Hedin}  L. Hedin, Phys. Rev. {\bf 139}, A796 (1965).
\bibitem{reviewOnida}  G. Onida, L. Reining, and A. Rubio, Rev. of Mod. Phys.
{\bf 74}, 601 (2002) and references therein.
\bibitem{bassani} F. Bassani and G. Pastori Parravicini,
"Electronic States and Optical Transitions in Solids", Pergamon
Press, New York 1975.
\bibitem{sym} The Si$_{10}$H$_{14}>$O cluster we considered here
corresponds to the Si$_{10}$H$_{14}$O-sym of \cite{gatti}, where O
make a bridge between "second neighbors" Si atoms. We have
obtained similar results have for the Si$_{10}$H$_{14}$O-asym
case, where O is in between two "first neighbors" Si atoms.
\bibitem{pwscf} The DFT calculations have been performed using the
ESPRESSO package: S. Baroni, A. Dal Corso, S. de Gironcoli, P.
Giannozzi, C. Cavazzoni, G. Ballabio, S. Scandolo, G. Chiarotti,
P. Focher, A. Pasquarello, K. Laasonen, A. Trave, R. Car, N.
Marzari, A. Kokalj, http://www.pwscf.org/.
\bibitem{del} C. Delerue, M. Lannoo, and G. Allan, Phys. Rev. Lett. {\bf 84}, 2457 (2000).
\bibitem{size} A careful discussion about the role of size on the
 many-body effects and of structural deformation on the
optical spectra will be presented in O. Pulci et al., to be
published.
\bibitem{Ma} Z. Ma, X. Liao, G. kong, J. Chu, Appl. Phys. Lett.
{\bf 75}, 1857 (1999).
\bibitem {daldo} N. Daldosso, M. Luppi, S. Ossicini, E. Degoli, R. Magri, G. Dalba, P.
Fornasini, R. Grisenti, F. Rocca, L. Pavesi, S. Boninelli, F.
Priolo, C. Spinella, and F. Iacona, Phys. Rev. B {\bf 68}, 085327
(2003).
\bibitem{dalne} L. Dal Negro, M. Cazzanelli, L. Pavesi, S. Ossicini, D. Pacifici, G. Franz\`o, F. Priolo,
and F. Iacona, Appl. Phys. Lett. {\bf 82}, 4636 (2003).
\bibitem{tersoff} Y. Tu, and J. Tersoff, Phys. Rev. Lett. {\bf 89}, 086102 (2002).
\bibitem{kelires} G. Hadjisavvas, and P. Kelires, Phys. Rev. Lett. {\bf 93}, 226104 (2004).
\end{thebibliography}
\end{document}